\documentclass[conference]{IEEEtran}
\IEEEoverridecommandlockouts
\usepackage[letterpaper, top=0.75in, bottom=1.037in, left=0.624in, right=0.624in]{geometry}
\usepackage[draft,bookmarks=false]{hyperref}
\usepackage{gensymb}
\usepackage{cite}
\usepackage{amsmath,amssymb,amsfonts}
\usepackage{algorithmic}
\usepackage{graphicx}
\usepackage{textcomp}
\usepackage{xcolor}
\usepackage[nameinlink]{cleveref}
\crefname{equation}{}{}
\crefname{section}{Sec.}{Secs.}
\crefname{figure}{Fig.}{Figs.}
\usepackage{balance}
\usepackage[lang=en]{jabbrv}
\makeatletter
\newcommand{\linebreakand}{%
  \end{@IEEEauthorhalign}
  \hfill\mbox{}\par
  \mbox{}\hfill\begin{@IEEEauthorhalign}
}
\makeatother
\def\BibTeX{{\rm B\kern-.05em{\sc i\kern-.025em b}\kern-.08em
    T\kern-.1667em\lower.7ex\hbox{E}\kern-.125emX}}
\ifCLASSOPTIONcompsoc
 \usepackage[caption=false,font=normalsize,labelfont=sf,textfont=sf]{subfig}
\else
 \usepackage[caption=false,font=footnotesize]{subfig}
\fi
\begin{document}
\bstctlcite{IEEEexample:BSTcontrol}

\title{RadAround: A Field-Expedient Direction Finder for Contested IoT Sensing \& EM Situational Awareness
\thanks{This project has been supported by the Embry-Riddle Aeronautical University Office of Undergraduate Research and Natalie \& Chris Maute.}
}

\author{\IEEEauthorblockN{Owen~Maute, Blake~Roberts and Berker~Peköz
}
\IEEEauthorblockA{
College of Engineering, Embry-Riddle Aeronautical University,
Daytona Beach, FL, USA\\
e-mail: \{\textit{\href{mailto:mauteo@my.erau.edu}
{mauteo}},\textit{\href{mailto:roberb25@my.erau.edu}
{roberb25}}\}\textit{@my.erau.edu}, \textit{\href{mailto:Berker.Pekoz@erau.edu}
{Berker.Pekoz@erau.edu}}}
}
\maketitle

\begin{abstract}
This paper presents RadAround, a passive 2-D direction-finding system designed for adversarial IoT  sensing in contested environments. Using mechanically steered narrow-beam antennas and field-deployable SCADA software, it generates high-resolution electromagnetic (EM) heatmaps using low-cost COTS or 3D-printed components. The microcontroller-deployable SCADA coordinates antenna positioning and SDR sampling in real time for resilient, on-site operation. Its modular design enables rapid adaptation for applications such as EMC testing in disaster-response deployments, battlefield spectrum monitoring, electronic intrusion detection, and tactical EM situational awareness (EMSA). Experiments show RadAround detecting computing machinery through walls, assessing utilization, and pinpointing EM interference (EMI) leakage sources from Faraday enclosures.


\end{abstract}


\section{Introduction}

Modern mission-critical IoT and sensing environments are saturated with RF emissions. While natural sources like lightning and cosmic rays contribute, most RF radiation comes from human-made systems: communications links for mobile terminals, unmanned aerial vehicles (UAVs), and radar platforms across land, sea, and air. Even unintended emissions from digital electronics, such as internal circuit switching, contribute to the ambient RF landscape. Some emissions originate from unauthorized or compromised devices, such as rogue communication modules in imported electronics \cite{mcfarlane_rogue_2025}.

Ambient RF signals support diverse sensing applications, including human motion \cite{9949562} and sleep posture \cite{10806892} classification, respiration monitoring \cite{10679710} and radiometric identification of humans through walls\cite{
10966040}. Beyond human-centric uses, RF emissions can be used to tomograph structures\cite{7944785} and reconstruct occluded visual scenes using reconfigurable intelligent surfaces (RISs) \cite{10907242} or even Wi-Fi-based RF inpainting \cite{9919801}. 



Though often benign, these emissions can be exploited by adversaries for passive reconnaissance, structural mapping, or situational awareness. Detecting, localizing, and characterizing them is mission-critical. RF mapping also validates low-noise satellite communication testing environments, identifies EMC violations in shielded enclosures to prevent unintended emissions on the battlefield, and optimizes wireless deployments.

Despite longstanding interest in EM situational awareness (EMSA) within this community, most solutions rely on spatially distributed sensor networks.
Examples of distributed sensor network monitoring include bearing angle tracking of HF transmitters  \cite{1180427}, 4-D cross-ambiguity function (CAF) derivations for co-located kinetic emitters\cite{7795452}, and Internet of Things (IoT) spectrum monitors \cite{8599695}, which are vulnerable to data falsification attacks attacks when used for Internet of Battlefield Things (IoBT)\cite{10356249}. Even four high-altitude ballons (HABs) can localize radios within a few hundred meters \cite{10773949}.

Beyond distributed sensor networks, there are low resolution and fragile solutions \cite{noauthor_wifi_nodate}. Infrared cameras are used to detect radiation-induced heating in certain ambient temperatures\cite{9899802}. 

Other approaches involve antenna arrays, dubbed microwave cameras in this context\cite{Microcam}. While electronically scanned arrays offer faster acquisition than mechanical scanning, they require calibration routines and either non-trivial beamforming ICs or multi-channel data acquisition (DAQ) pipelines; making them impractical for rapid field-expedient deployment. Among such systems most aligned to this work's sensitivity, spatial resolution, and operational simplicity is \cite{10739065}; however, it requires precise geometric and circuit-level construction involving multiple PCBs populated with a variety of surface-mounted components, limiting field reproducibility. Its hardware configuration, specifically the use of ESP32s, restricts operation to 2.4~GHz ISM band. Comparable systems described in \cite{8485021,9245683} depend on commercial dosimeters and Kinect sensors, or magnetic field meters, further constraining their deployability.

 
 
 Another practical limitation is that such EMSA systems, when available for the parameters of interest, are usually subject to Export Administration (EAR) and International Traffic in Arms Regulations (ITAR), which hinders their field-ready  deployment to operational environments
 . This underscores  the need for a high-fidelity direction finding system that can be rapidly manufactured with minimal components available in the field, tailored to the custom parameters of interest.

To address these operational gaps, this paper presents RadAround: a direction finder capable of resilient IoT device sensing, designed for rapid deployment in contested, austere or denied environments.  It investigates how low-cost commercial off-the-shelf (COTS) components that are either easily sourced or already available in the field, and a 3-D printer can be used to rapidly design and deploy a direction finding system, and evaluates its performance. Unlike traditional systems that rely on complex antenna arrays or distributed sensor networks, RadAround employs a single mechanically steered, narrow-beam antenna paired with a software-defined radio (SDR) and a lightweight, microcontroller-compatible supervisory control and data acquisition (SCADA) framework that runs offline.

RadAround enables high-resolution EM imaging across user-defined frequency bands, area and spatial resolutions. It can be configured, built, and deployed within a day, making it suitable for time-sensitive missions such as EMC diagnostics, tactical spectrum monitoring, and interference localization. 

This paper investigates the feasibility and performance of RadAround through a series of experimental campaigns, including detection of computing machinery through walls, assessment of processor utilization levels via monitoring covert RF signatures, and identification of EM leakage from Faraday enclosures. The results demonstrate that even a proof-of-concept implementation of RadAround can deliver actionable EMSA with minimal infrastructure and operator training.
This distinguishes it from state-of-the-art EM field (EMF) sensors, which employ Nintendo WII remotes and cameras to detect EM interference (EMI) from sources such as lamps \cite{8485021} and anti-theft sensors \cite{9245683}, of which visualization resolutions were constrained to $18 \times 12$ pixels, and $11\times20$ pixels, respectively, over several meters. This work contributes a novel integration of mechanically steered narrowbeam antennas with microcontroller-driven SCADA for high-resolution electromagnetic imaging using only field-accessible components.

The rest of this work is organized as follows: \Cref{sec:rfsys,sec:mech} present the rapidly deployable RF \& mechanical systems, respectively; \Cref{sec:scada} discusses the rationale of the SCADA system, \Cref{sec:exp} explores the experimental verification campaign, and finally \Cref{sec:conc} highlights conclusions and discusses findings as well as future extensions.

\section{RF System Architecture and Construction\label{sec:rfsys}}
To minimize front-end complexity and ensure compatibility with single-port SDRs, RadAround employs a single highly directive antenna rather than a phased array, as used in prior art\cite{10739065}. This design choice reduces system cost, simplifies integration, and enhances field adaptability.

\subsection{Narrowbeam 2.4~GHz ISM Monitor\label{narrowbeamant}}
A rugged left-hand circularly polarized (LHCP) helicone antenna was designed to monitor the 2.4~GHz ISM band. The antenna was modeled using multiple CAD tools and fabricated via 3-D printing, with a steel mesh conical reflector added to improve directivity and suppress sidelobes. The helicone architecture was selected for its lightweight, modular construction and high gain across all polarizations except RHCP, making it ideal for passive sensing of diverse emitters.

\begin{figure}[htbp]
    \centering
    \includegraphics[trim=2.50in 1.37in 1.17in 1.42in, clip, angle=90, width=.96\columnwidth]{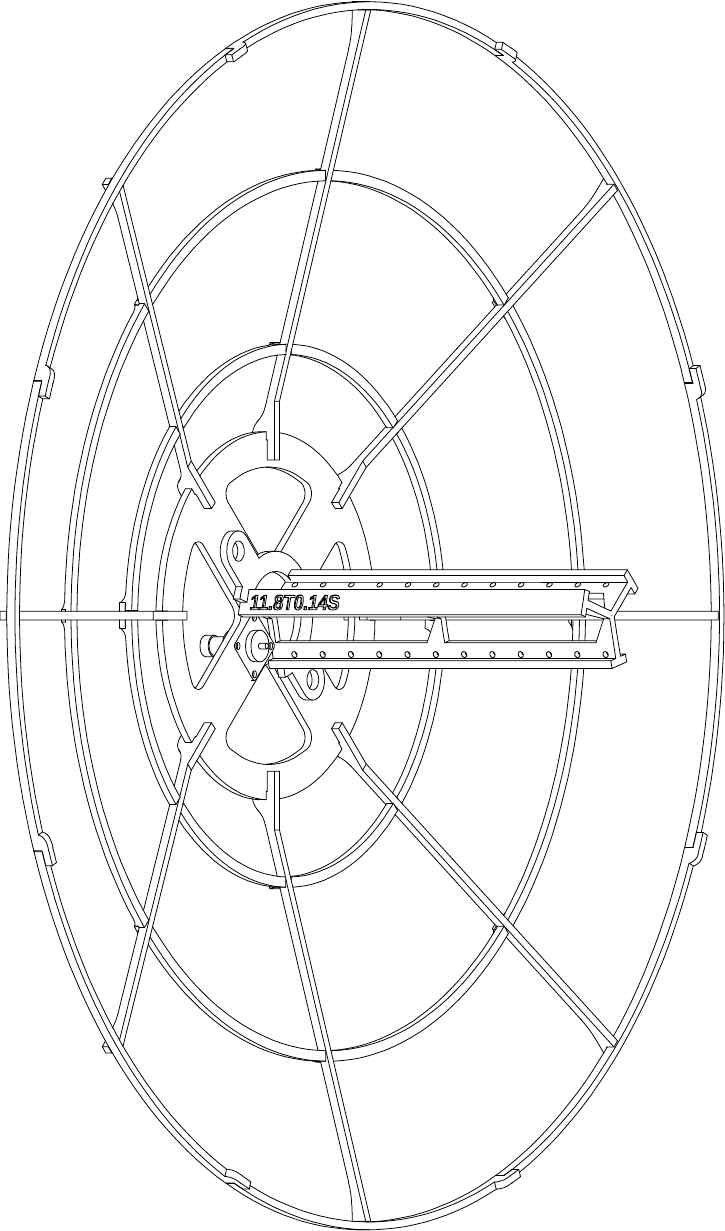}
    \caption{3-D rendering of the final 2.4~GHz ISM antenna structure, including the helical scaffold and the conical reflector.}
    \label{fig:wifiant}
\end{figure}

The helical scaffold was modeled  off of \cite{dsgc_helix_v5}. To tune the antenna to 2.45 GHz while maintaining compatibility with 3-D printer capabilities, the design parameters were modified to 13 turns, 12~AWG wire radius, and an $11.3\degree$ pitch. The resulting design achieved a simulated gain of 14.9~dBi \cite{kraus_antennas_2002,helix_calc_tool}, prior to reflector integration. The antenna design was parameterized in OpenSCAD\cite{openscad}, enabling rapid reconfiguration on resource-limited systems without the need for graphical interfaces, an important feature for field-deployable systems. The reflector geometry was based on guidelines from \cite{4584946} and modeled in Autodesk Fusion 360. A 3-D rendering of the complete structure is presented in \cref{fig:wifiant}.

The entire assembly comprising the helical scaffold and reflector was printed using PETG filament \cite{prusament_petg} on a Prusa MK1 3-D printer \cite{prusa_mk1}, and assembled with M3 screws to ensure mechanical robustness and proper grounding. An SMA connector was mounted on the baseplate to interface with RF measurement equipment.
A stainless steel mesh sheet with a wire diameter of 0.20~mm and square weave openings of 0.43~mm, as would be available in the field, was applied to the conical surface to enhance its EM reflectivity. 

\subsection{High-gain Wideband Electronics Finder at  1.7~GHz Band}
To detect covert RF signatures from computing and communication systems, RadAround includes a second RF system tuned to the 1.7~GHz band. Normally allocated to mobile satellite services (1.6~GHz) and broadband communications (1.7-1.8~GHz), this frequency range is of particular interest due to EMI from common digital subsystems, including:

\begin{itemize}
    \item Modern computer and smartphone CPUs often operate at core frequencies around 3.4~GHz, derived by multiplying low frequency base clocks (e.g., 25, 40, or 100~MHz) through phase-locked loops (PLLs). Penultimate frequencies around 1.7~GHz appear as spurious emissions.
    \item Router CPUs frequently operate directly at or near 1.7~GHz, contributing to emissions in that band.
    \item DDR3 memory interfaces are typically clocked at 1600 or 1866~MHz. Newer  generations use the same as internal reference clocks, producing EMI near 1.7~GHz.
    \item PCI Express (PCIe) systems use a 100~MHz reference clock, and present harmonics near 1.6~GHz as reference clock is multiplied, contributing to EMI in those bands.
\end{itemize}

Since compute, memory and interface components emit little switching-induced EMI, and we observe a harmonic rather than the ultimate switching frequency in most cases, a low noise amplifier (LNA) is introduced to the system design, in addition to the antenna, to enhance detection. As this work focuses on constructing desired system with COTS components that are readily accessible in the field or widely distributed, a NooElec SAWbird+ GOES module \cite{sawbird_goes} is integrated. This module is specifically designed to receive GOES weather satellite downlinks typically needed in the field. It is therefore centered at 1688~MHz, offers 38~dB of gain, an 80~MHz 3~dB bandwidth, and a 1.2~dB noise figure.

To simulate rapid deployment, the team adapted a publicly available 3-D printable RHCP helicone designed for NOAA satellite reception\cite{thing6436342} by eliminating non-essential structural elements surrounding the reflector to reduce weight \& simplify geometry for expedited fabrication. The cone was covered with the same stainless steel mesh material used in \cref{narrowbeamant}.

\section{Mechanical System\label{sec:mech}}
To enable full-environment scanning using a single highly directional antenna, RadAround employs a mechanically actuated beam-steering platform. This two-degree-of-freedom (2DOF) rotor system allows for precise angular control in both azimuth and elevation, enabling spherical coverage.

For azimuthal rotation, the system uses a large herringbone gear with a 300:12 gear ratio. This configuration enhances angular resolution and torque, ensuring smooth and accurate positioning even under load. Elevation control is achieved via a worm gear mechanism, chosen for its inherent resistance to backdriving, an essential feature for maintaining stability when supporting the weight of the antenna and reflector assembly.

\begin{figure}[htbp]
    \centering
    \includegraphics[trim=2.3in 4.58in 1.28in 3.04in, clip, width=1\columnwidth]{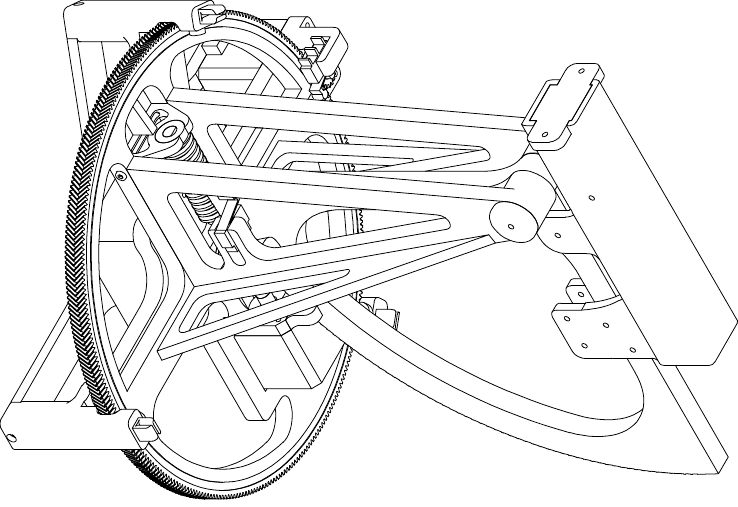}
    \caption{3-D CAD rendering of the mechanical system in Fusion 360 (rotated quarter turn clockwise to save space).}
    \label{fig:rotor}
\end{figure}

\begin{figure}[htbp]
    \centering
    \includegraphics[ width=.95\columnwidth]{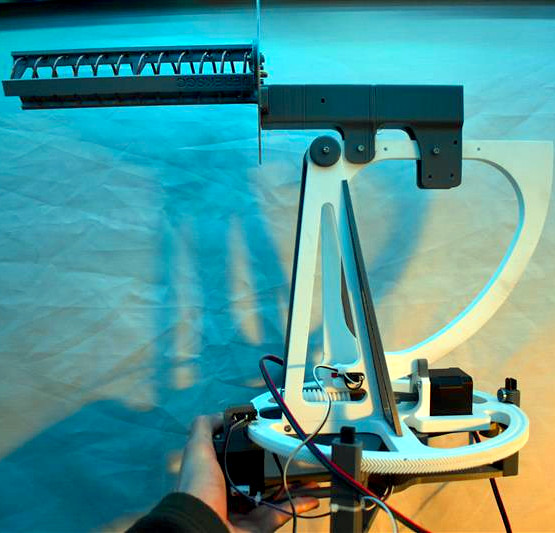}
    \caption{Photo of manufactured RF and mechanical system components.}
    \label{fig:photo}
\end{figure}

The antenna is housed in a compact, hot-swappable mount that minimizes the system’s footprint and allows for rapid interchangeability between frequency-specific front-ends. This modularity supports mission-specific reconfiguration in the field without requiring specialized tools or recalibration. The complete rotor assembly is drawn in \cref{fig:rotor} and photographed in \cref{fig:photo}. All components were fabricated using the process described in \cref{sec:rfsys}, leveraging PETG filament for its balance of strength and flexibility.

To demonstrate cost efficiency, the stepper motors used for actuation were salvaged from high-volume office printers, specifically from paper feed and tray lift mechanisms. This reuse of available components aligns with the system’s design philosophy of leveraging field-accessible hardware to enable field-ready deployment in austere environments.

\section{Supervisory Control and Data Acquisition\label{sec:scada}}
 The rotor is required to operate in user-defined elevation and azimuth range intervals. A microcontroller that communicates with a process supervisor, controls the stepper motors and interfaces with limit switches is thus needed to perform homing procedures and ensure accurate and timely positioning. The signals from the RF front-end cannot be acquired without a radio. A supervisor that commands the microcontroller and radio harmoniuously and collects and processes data is needed.
 
The SCADA software was developed in MATLAB due to its robust support for hardware interfacing, signal processing, and real-time control. MATLAB’s cross-platform compatibility and ability to compile scripts for deployment on embedded systems make it well-suited for field-adaptable applications. The software interfaces with both the microcontroller and the SDR, issuing movement commands, managing frequency hopping, and aggregating received signal data.

The SDR was selected with field adaptability in mind. As HackRF One is the cornerstone SDR in the field due to its PortaPack \cite{portapack_h4m} hardware add-on, and compatible with the SCADA suite through manufacturer provided cross-platform drivers, it was used in the prototype design. The authors note that ADALM-PLUTOs are more cost effective and better suited for this specific task with their wider bandwidth and dynamic range, the results demonstrated in this work establish a   baseline that would be improved by using ADALM-PLUTO.

The microcontroller platform is an Arduino Uno, chosen for its real-time operating system (RTOS) as a microcontroller, and rapid integration of stepper motors and limit switches through CNC motor shield, enabling precise control of the azimuth and elevation axes. Although a Raspberry Pi (RPi) could support a more embedded and self-contained implementation, the Arduino-based approach was prioritized to minimize development time and complexity under resource-constrained conditions to align with this work.

\begin{figure}[htbp]
    \centering
    \includegraphics[width=.88\columnwidth]{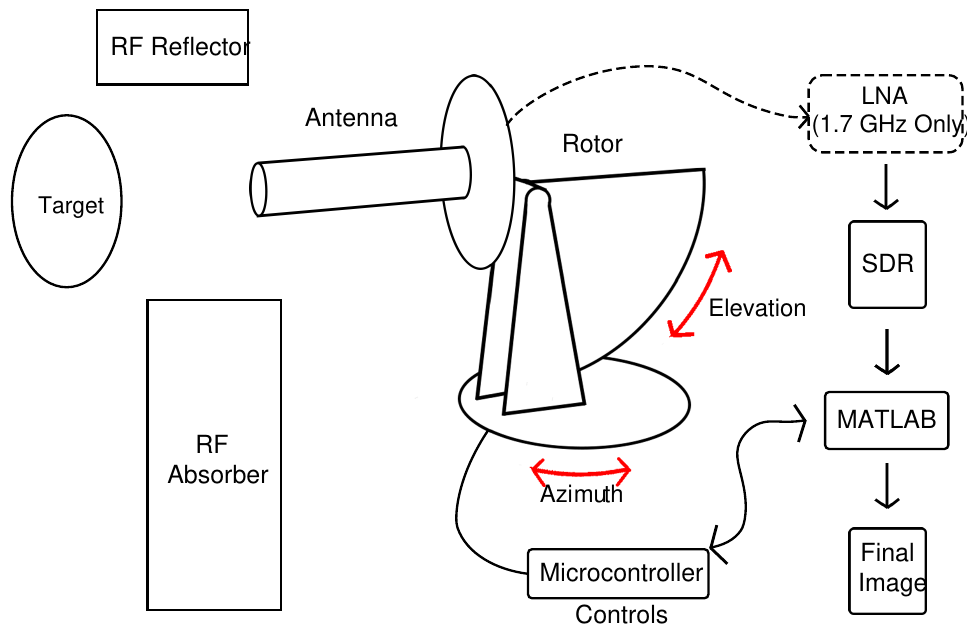}
    \caption{Block diagram of the system in the environment.}
    \label{fig:intsys}
\end{figure}

The integrated system is illustrated in \cref{fig:intsys}. 
The SCADA system operates as follows: the host computer communicates with the microcontroller via serial-over-USB and with the SDR via USB. Upon initialization, the SCADA software homes the rotor using limit switches, then begins scanning by stepping through user-defined azimuth and elevation angles. At each angular position, the SDR captures IQ samples across a sequence of frequency centers. A secondary thread computes the power of each sample set and updates the corresponding pixel in the EM heatmap. Azimuthal and frequency sweeps alternate directions between elevation and azimuth steps, respectively, to optimize acquisition time and minimize mechanical wear.


\section{Experimental Verification Campaign\label{sec:exp}}

\subsection{Antenna characterization}
\begin{figure}[htbp]
    \centering
    \includegraphics[trim=6.3in 11.58in 1.28in 9.04in, clip, width=.94\columnwidth]{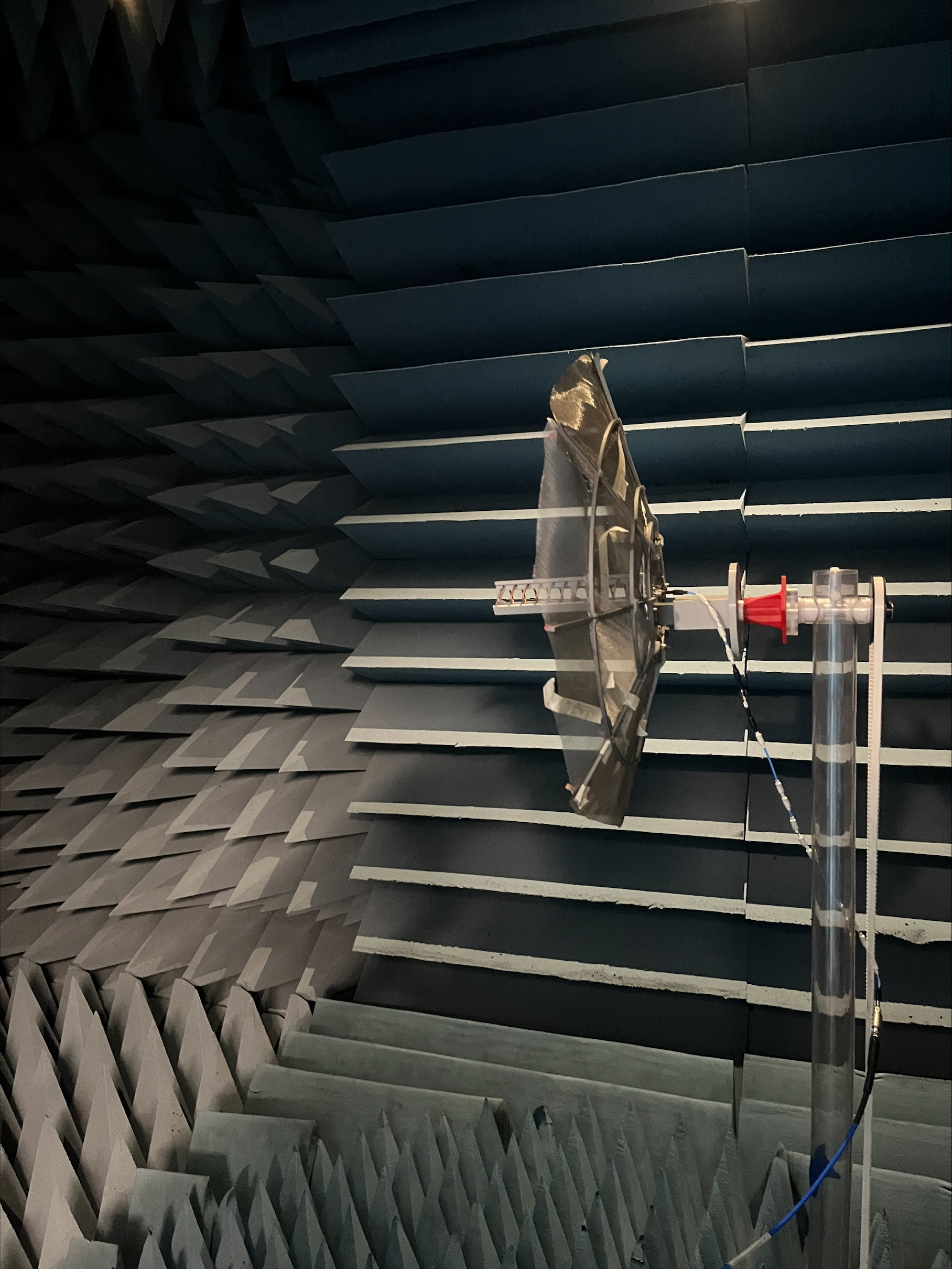} 
    \caption{2450 MHz antenna shown inside far field chamber.}
    \label{fig:setUp}
\end{figure}

The two antennas were tested in an ETS Lindgren far field anechoic chamber using a Keysight Vector Network Analyzer (VNA). 
Calibration was performed using a reference antenna to compensate for system losses and to determine the gain of the fixed receive antennas, following the methodology in\cite{BalanisConstantineA2016AM}. Measurements were conducted for both horizontal and vertical polarizations, enabling derivation of circular polarization patterns via ETS EMQuest software.  One antenna is then placed in the anechoic chamber as shown in \cref{fig:setUp}. 

\begin{figure}[htbp]
    \centering
    \subfloat[\label{fig:c}]{
        \centering
        \includegraphics[trim=2.23in 3.32in 2.1in 3.32in, clip, width=0.45\columnwidth]{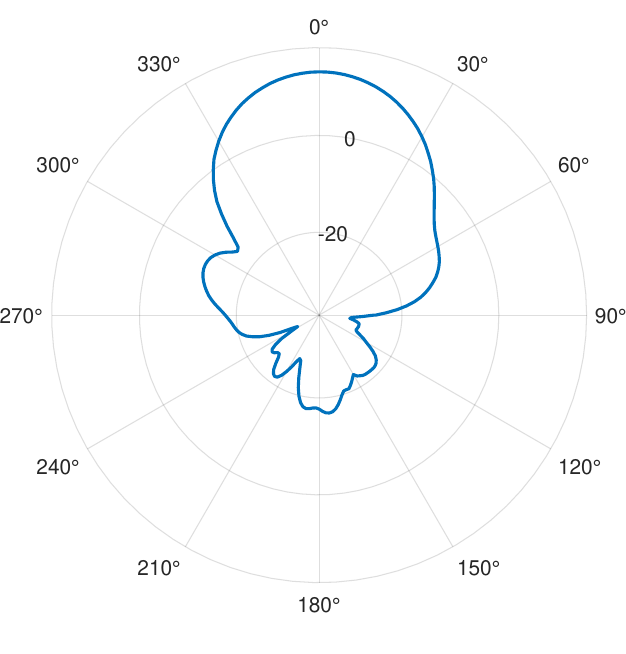}}
    \hfill
    \subfloat[\label{fig:d}]{
        \centering
        \includegraphics[trim=2.23in 3.32in 2.1in 3.32in, clip, width=0.45\columnwidth]{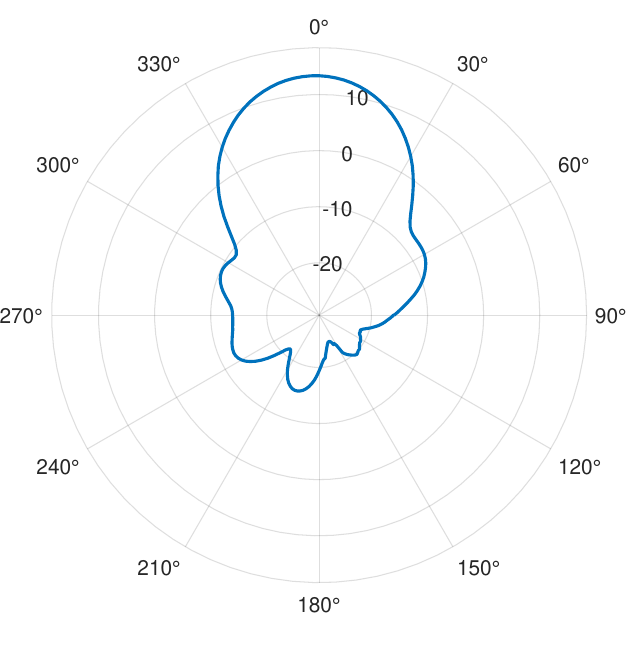}}
        
    \subfloat[\label{fig:a}]{
        \centering
        \includegraphics[trim=2.23in 3.32in 2.1in 3.32in, clip, width=0.45\columnwidth]{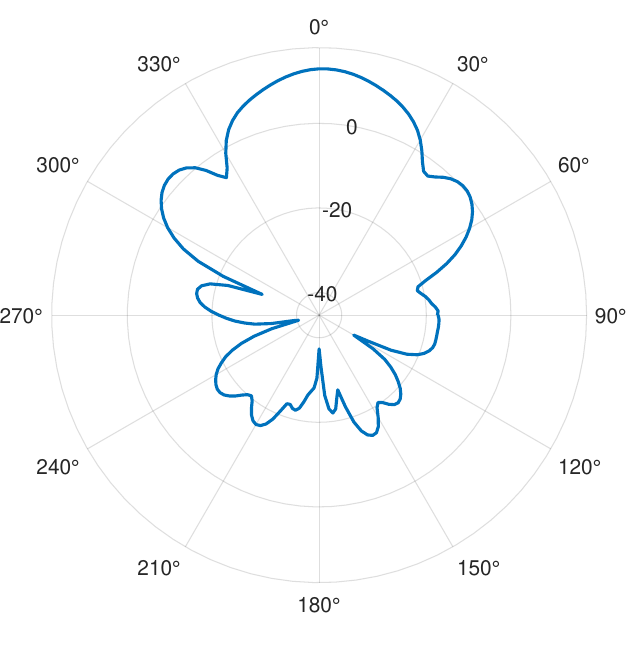}}
    \hfill
    \subfloat[\label{fig:b}]{
        \centering
        \includegraphics[trim=2.23in 3.32in 2.1in 3.32in, clip, width=0.45\columnwidth]{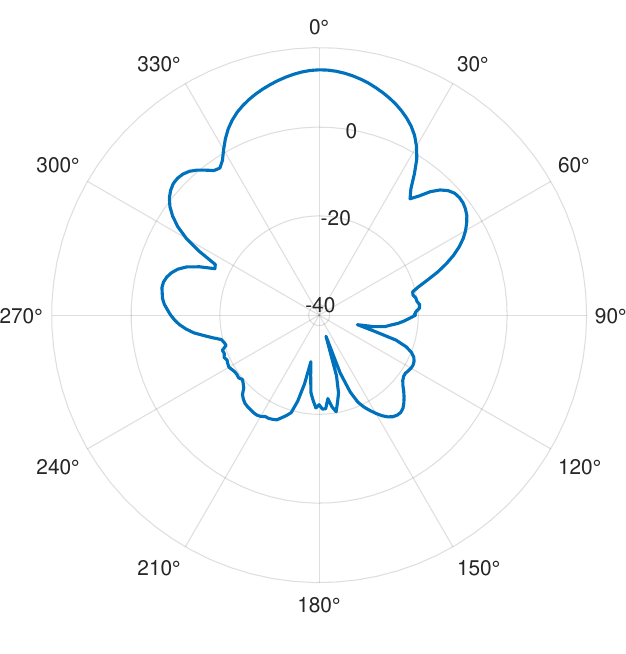}}

    \caption{Realized gain patterns for antennas: 1700~MHz (a) $\phi$ and (b) $\theta$ planes, 2450~MHz (c) $\phi$ and (d) $\theta$ planes.}
    \label{fig:pattern}
\end{figure}

\begin{figure}[htbp]
    \centering
    \subfloat[\label{fig:1750bw}]{
        \centering
        \includegraphics[width=0.48\columnwidth]{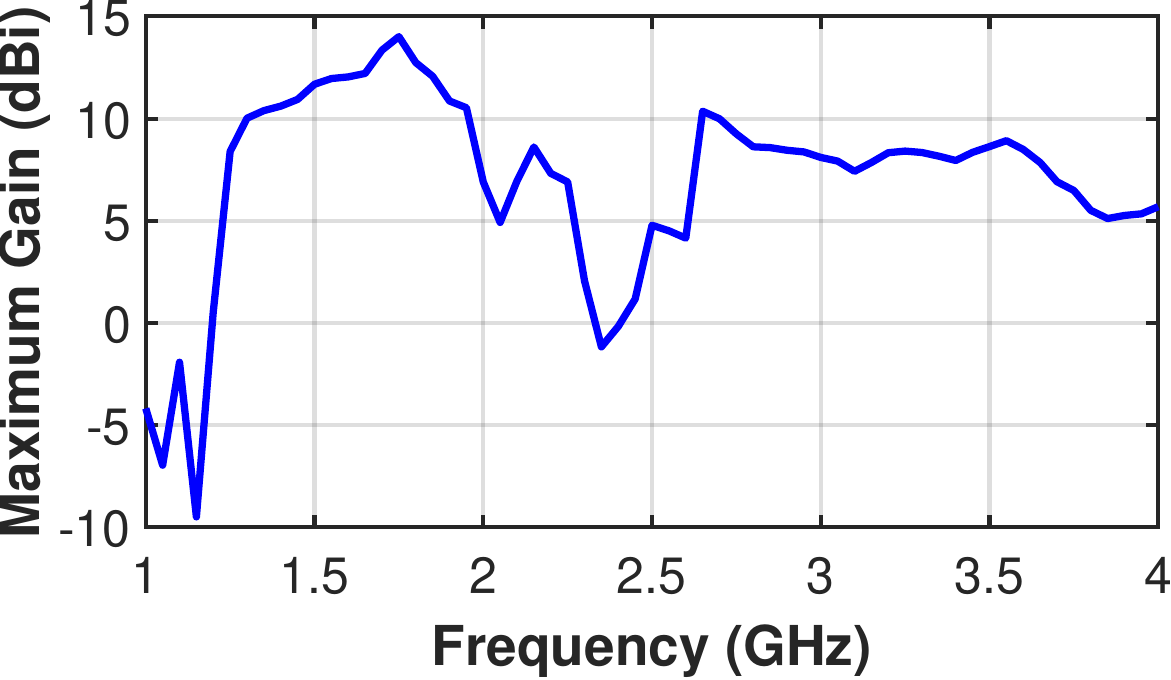}}
\hfill
    \subfloat[\label{fig:2400bw}]{
        \centering
        \includegraphics[width=0.48\columnwidth]{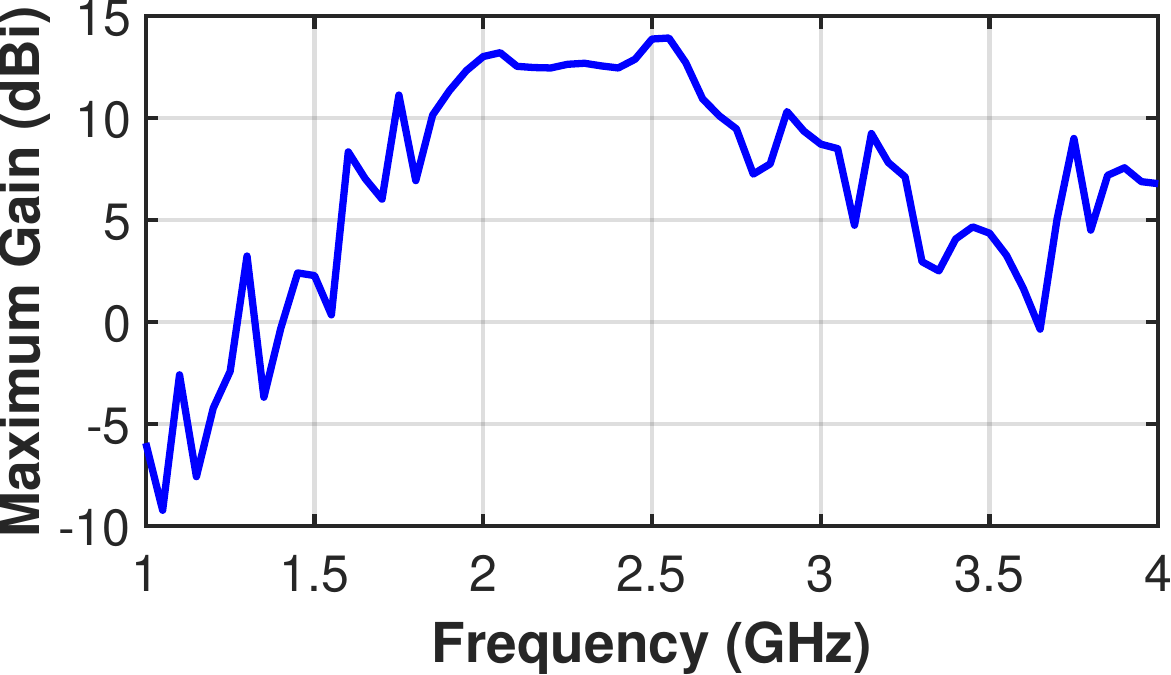}}
    \caption{Operational bandwidth of (a) 1.7~GHz and (b) 2.45~GHz antennas.}
    \label{fig:opbw}
\end{figure}

The antennas were rotated  $360\degree$
in  $2\degree$ increments, and the resulting data were exported and visualized in MATLAB  as seen in \cref{fig:pattern}. A slice of the full pattern is shown along the $\phi$ plane ($\theta=0$) and the $\theta$ plane ($\phi=0$).
The 1.7~GHz antenna exhibited a boresight gain of 14.01~dBi with a 3~dB beamwidth of approximately 40\degree, while the 2.45~GHz antenna achieved a boresight gain of 13.94~dBi and a 3~dB beamwidth of approximately 30\degree. The operational bandwidths, shown in \cref{fig:opbw}, span 350~MHz (1.5–1.85~GHz) for the 1.7~GHz antenna and 750~MHz (1.9–2.65~GHz) for the 2.4~GHz antenna.

\subsection{Dwell and Integration Times}

The dwell time per angular step had to be set to a minimum of 0.5~seconds to mitigate mechanical vibrations ensuring measurement fidelity. The total acquisition time for a $100\times100$~pixel scan covering $180\degree$  azimuth and $80\degree$  elevation was  $\approx 1.4$~hours. The SDR’s 20~MHz bandwidth necessitated frequency sweeping: the 1.7~GHz scan integrated four 0.125-second measurements across the 80~MHz bandwidth centered at 1688~MHz, while the 2.4~GHz scan integrated five 0.1-second measurements across the 2.4–2.5~GHz ISM band.

\subsection{Data Collection Scenarios}

\subsubsection{Detecting Computing Systems at 1.7~GHz}
\begin{figure}[htbp]
    \centering
    \includegraphics[trim=0 0 0 7em, clip, width=\columnwidth]{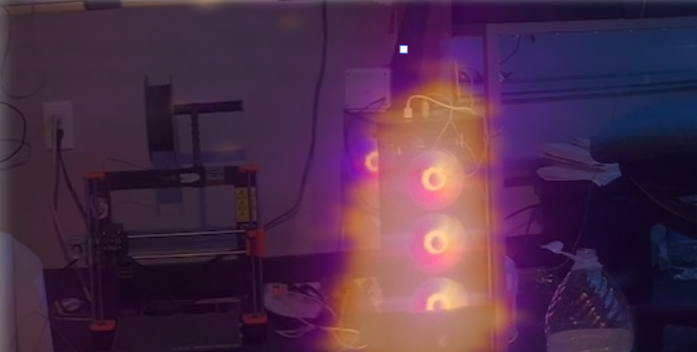} 
    \caption{1.7~GHz measurement of a desktop computer with a tempered glass case operating at 100\% CPU utilization, overlaid over optical photograph at boresight of RadAround.}
    \label{fig:overlay}
\end{figure}

RadAround was positioned 6~feet from a desktop computer operating at 100\% CPU utilization. EM measurement was laid over optical imagery capturing a $90\degree$ horizontal and $30\degree$ vertical field of view. The results  in \cref{fig:overlay} confirm the system’s ability to localize computing devices based on covert RF signatures.

\subsubsection{Computing Utilization Level Discrimination at 1.7~GHz}
\begin{figure}[htbp]
    \centering
    \includegraphics[width=\columnwidth]{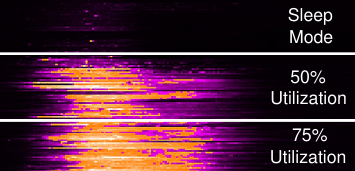} 
    \caption{1.7~GHz measurement of a desktop CPU operating at sleep mode, 50\% and 75\% CPU utilization (clipped all scans to contents of 75\%).}
    \label{fig:cpuutil}
\end{figure}

The same desktop CPU was operated at varying utilization levels (sleep, 50\%, 75\%). The system captured the results shown in  \cref{fig:cpuutil}, detecting EMI variations corresponding to utilization, with increased intensity and spatial spread at higher loads. Even in sleep mode, residual emissions were visible, indicating detectable persistent subsystem activity.


\subsubsection{Through-Wall Electronics Detection at 1.7~GHz}

\begin{figure}[htbp]
    \centering
    \includegraphics[trim=0 1ex 0 1em, clip,width=.87\columnwidth]{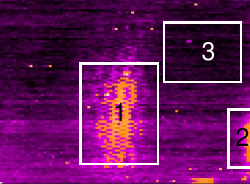} 
    \caption{Computing machinery detection through the wall. 1 is a running laptop, 2 is reflections off of HVAC plate, 3 is potential insulation material.}
    \label{fig:hello}
\end{figure}

RadAround was placed 4 feet away from a wooden door, behind which a consumer laptop operated at 50\% CPU utilization during the scan. \cref{fig:hello} shows the system successfully detected the laptop and secondary reflections from heating, ventilation, and air conditioning (HVAC) infrastructure and insulation materials, demonstrating through-wall sensing capability.

\subsubsection{Wi-Fi Access Point (AP) Localization Through Walls and Environmental Mapping at 2.4~GHz}

\begin{figure}[htbp]
    \centering
    \includegraphics[trim=0 0 0 1em, clip,width=\columnwidth]{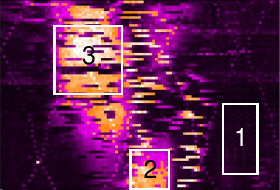} 
    \caption{2.4~GHz measurement of an AP behind a wall. Region 3 indicates the AP location, Region 2 corresponds to an HVAC vent. The bright regions between 1 and 2 are reflections from the corner of a wall perpendicular to the one obscuring the AP. Region 1 denotes the wall of the adjacent room.}
    \label{fig:aploc}
\end{figure}

A 2.45~GHz scan was conducted with the system positioned 15~feet from a Wi-Fi AP, separated by a wall and door. While the AP remained clearly identifiable, signal bleeding increased at this distance, consistent with prior art \cite{8485021,9245683,10739065}. As transmitter-receiver separation increases, multipath propagation becomes more prominent, leading to greater delay spread and spatial ambiguity due to reflected paths dominating over the direct signal, particularly in cluttered indoor scenes with structural features such as HVAC vents and wall boundaries. Diffraction artifacts originating from the 2.4~GHz antenna's reflector support structure were also evident in \cref{fig:aploc}, aligning with prior diffraction analyses\cite{schmidt1973diffraction}.  These results demonstrate that the system enables not only radiator localization, but also passive environmental mapping through reflection analysis.

\section{Conclusion\label{sec:conc}}


This paper introduced RaDAround, a modular, rapidly field deployable direction-finding system constructed entirely from low-cost COTS components and 3D-printed structures. The system demonstrated the ability to generate high-resolution EM imagery at user-defined frequencies and angular resolutions. Experimental validation confirmed its capability to detect computing systems through walls, assess processor activity levels, and localize RF sources such as Wi-Fi access points. 

RaDAround’s ability to detect unintentional EMI and covert RF signatures; and localize RF sources supports adversarial spectrum dominance and battlefield IoT resilience. Its low size, weight, power, and cost (SWaP-C) profile, combined with field-adaptable design, makes it suitable for harsh, adversarial, austere or denied environments. Future enhancements include:

\begin{itemize}
    \item Rotor stabilizing vibration mitigation reducing dwell time,
    \item Employing wider-bandwidth SDRs matching antennas to improve spectral coverage and resolution,
    \item 3D triangulation using two scanning antennas,
    \item Embedding SCADA functionality on microcontrollers to enable standalone operation without a host computer,
    \item Integrating printer-sourced motor drivers to further align with austere deployment constraints,
    \item Machine learning to distinguish radiators from multipath reflectors, leverage multipath artifacts for passive environmental mapping, classify computing systems based on RF signatures, estimate utilization levels and distances.
\end{itemize}
These improvements aim to transition RaDAround from a proof-of-concept to a field-ready system aligned with operational needs in RF sensing and tactical spectrum dominance.

\section*{Acknowledgments}

The authors thank Dr. Alphan Şahin for his help identifying reflector structure diffraction artifacts; Nathan Barnes, and Gabe Emerson for their valuable insights that contributed to early development and understanding, and Cameron Martinez for helping obtain \cref{fig:pattern,fig:opbw}.
Portions of this manuscript were augmented with the assistance of Microsoft 365 Copilot Researcher and Writing Coach Agents (Microsoft, 2025). The final content was reviewed and confirmed by the authors.

\bibliographystyle{jabbrv_IEEEtran}
\balance
\bibliography{references}

@IEEEtranBSTCTL{IEEEexample:BSTcontrol,
  CTLuse_forced_etal = "yes",
  CTLmax_names_forced_etal = "5",
  CTLnames_show_etal = "3",       
}

@INPROCEEDINGS{9899802,
  author={Mahalec, Rene and Malarić, Krešimir},
  booktitle={Proc. 2022 Int. Symp. ELMAR}, 
  title={Assessment of Microwave Radiation Using Thermal Cameras}, 
  year={2022},
  pages={19-22},
  doi={10.1109/ELMAR55880.2022.9899802}
}

@ARTICLE{9919801,
  author={Chen, Cheng and Nishio, Takayuki and Bennis, Mehdi and Park, Jihong},
  journal={IEEE Access}, 
  title={{RF}-Inpainter: Multimodal Image Inpainting Based on Vision and Radio Signals}, 
  year={2022},
  volume={10},
  number={},
  pages={110689-110700},
  doi={10.1109/ACCESS.2022.3214972}}

@misc{noauthor_wifi_nodate,
  title = {{WiFi} Camera},
  author = {Emaad Paracha and Ton, Andrew},
  url = {https://wifiimagingcamera.com/index.html},
  year = {2014},
  urldate = {2025-05-13}
}

@INPROCEEDINGS{9949562,
  author={Yu, Cong and Zhang, Dongheng and Xie, Chunyang and Lu, Zhi and Hu, Yang and Li, Houqiang and Sun, Qibin and Chen, Yan},
  booktitle={Proc. 2022 IEEE 24th Int. Workshop on Multimedia Signal Process. (MMSP)}, 
  title={{WiFi}-Based Human Pose Image Generation}, 
  year={2022},
  pages={1-6},
  doi={10.1109/MMSP55362.2022.9949562}
}

@INPROCEEDINGS{10739065,
  author={Euchner, Florian and Brink, Stephan Ten},
  booktitle={Proc. 2024 Kleinheubach Conf.}, 
  title={{ESPARGOS}: Phase-Coherent {WiFi} {CSI} Datasets for Wireless Sensing Research}, 
  year={2024},
  pages={1-4},
  doi={10.23919/IEEECONF64570.2024.10739065}
}

@INPROCEEDINGS{1180427,
  author={Raps, F. and Kollmann, K. and Zeidler, H.C.},
  booktitle={Proc. 2002  IEEE Mil. Commun. Conf. (MILCOM)}, 
  title={Feature extraction for {HF}-band emitter location}, 
  year={2002},
  volume={1},
  pages={131-135},
  doi={10.1109/MILCOM.2002.1180427}
}

@INPROCEEDINGS{9245683,
  author={Sato, Ken and Kamimura, Yoshitsugu},
  booktitle={Proc. 2020 Int. Symp. on Electromagn. Compat. (EMC EUROPE)}, 
  title={A Simple Measurement Method of Electromagnetic Field Distribution using Machine-Learning}, 
  year={2020},
  pages={1-4},
  doi={10.1109/EMCEUROPE48519.2020.9245683}
}

@INPROCEEDINGS{8485021,
  author={Sato, Ken and Tsukahara, Tomoya and Kamimura, Yoshitsugu},
  booktitle={Proc. 2018 Int. Symp. on Electromagn. Compat. (EMC EUROPE)}, 
  title={Visualization of Electromagnetic Field Distribution with Augmented Reality}, 
  year={2018},
  pages={506-509},
  doi={10.1109/EMCEurope.2018.8485021}
}

@INPROCEEDINGS{Microcam,
  author={Yan, Xin and Liu, Liang and Khilkevich, Victor},
  booktitle={Proc. 2023 IEEE Symp. on Electromagn. Compat. \& Signal/Power Integrity (EMC+SIPI)}, 
  title={A Real-time Microwave Camera Prototype with Zero-bias Diode Detectors for {EMI} Source Imaging}, 
  year={2023},
  pages={614-618},
  doi={10.1109/EMCSIPI50001.2023.10241444}
}

@ARTICLE{10806892,
  author={Xu, Leiyang and Zheng, Xiaolong and Du, Xinrun and Liu, Liang and Ma, Huadong},
  journal={IEEE Transactions on Mobile Computing}, 
  title={{WiCamera}: Vortex Electromagnetic Wave-Based {WiFi} Imaging}, 
  year={2025},
  volume={24},
  number={5},
  pages={3633-3649},
  doi={10.1109/TMC.2024.3519623}}

@ARTICLE{10966040,
  author={Su, Can and Xue, Xinlei and Ma, Lei and Zhang, Xiaolong and Yan, Wei and Bian, Kaigui},
  journal={IEEE Internet of Things Journal}, 
  title={Robust Indoor Person Re-Identification with Multimodal Training}, 
  year={2025},
  volume={},
  number={},
  pages={1-1},
  doi={10.1109/JIOT.2025.3561213}}

@INPROCEEDINGS{8599695,
  author={Cohen, Aaron E. and Jiang, Gina G. and Heide, David A. and Pellegrini, Vincenzo and Suri, Niranjan},
  booktitle={Proc. 2018 IEEE Mil. Commun. Conf. (MILCOM)}, 
  title={Radio Frequency {IoT} Sensors in Military Operations in a Smart City}, 
  year={2018},
  volume={},
  number={},
  pages={763-767},
  doi={10.1109/MILCOM.2018.8599695}}

@INPROCEEDINGS{10356249,
  author={Sánchez Sánchez, Pedro Miguel and Tomás Martínez Beltrán, Enrique and Celdrán, Alberto Huertas and Wassink, Robin and Bovet, Gérôme and Pérez, Gregorio Martínez and Stiller, Burkhard},
  booktitle={Proc. 2023 IEEE Mil. Commun. Conf. (MILCOM)}, 
  title={Stealth Spectrum Sensing Data Falsification Attacks Affecting {IoT} Spectrum Monitors on the Battlefield}, 
  year={2023},
  volume={},
  number={},
  pages={673-678},
  doi={10.1109/MILCOM58377.2023.10356249}}

@INPROCEEDINGS{7795452,
  author={Watson, William and McElwain, Thomas},
  booktitle={Proc. 2016 IEEE Mil. Commun. Conf. (MILCOM)}, 
  title={{4D} {CAF} for localization of co-located, moving, and {RF} coincident emitters}, 
  year={2016},
  volume={},
  number={},
  pages={948-951},
  doi={10.1109/MILCOM.2016.7795452}}

@INPROCEEDINGS{10773949,
  author={Schäfer, Matthias and Lizarribar, Yago and Bovet, Gérôme and Verbruggen, Dieter},
  booktitle={Proc. 2024 IEEE Mil. Commun. Conf. (MILCOM)}, 
  title={Let’s Take This Upstairs: Localizing Ground Transmitters With High-Altitude Balloons}, 
  year={2024},
  volume={},
  number={},
  pages={475-480},
  doi={10.1109/MILCOM61039.2024.10773949}}

@ARTICLE{10679710,
  author={He, Qiuye and Yang, Edwin and Fang, Song and Zhao, Shangqing},
  journal={IEEE/ACM Transactions on Networking}, 
  title={Revisiting Wireless Breath and Crowd Inference Attacks With Defensive Deception}, 
  year={2024},
  volume={32},
  number={6},
  pages={4976-4988},
  doi={10.1109/TNET.2024.3453903}}

@INPROCEEDINGS{10907242,
  author={Zirak, Kavian and Imani, Mohammadreza F.},
  booktitle={Proc. 2025 US Nat. Comm. URSI Nat. Radio Sci. Meet. (USNC-URSI NRSM)}, 
  title={A Novel Computational Imaging Method Using Reconfigurable Intelligent Surfaces}, 
  year={2025},
  volume={},
  number={},
  pages={381-382},
  doi={10.23919/USNC-URSINRSM66067.2025.10907242}}

@INPROCEEDINGS{7944785,
  author={Karanam, Chitra R. and Mostofi, Yasamin},
  booktitle={Proc. 2017 16th ACM/IEEE Int. Conf. Inf. Process. Sensor Netw. (IPSN)
}, 
  title={{3D} Through-Wall Imaging with Unmanned Aerial Vehicles Using {WiFi}}, 
  year={2017},
  volume={},
  number={},
  pages={131-142},
  doi={}}

@article{mcfarlane_rogue_2025,
  author       = {Mcfarlane, Sarah},
  title        = {Ghost Machine: Rogue Communication Devices Found in Chinese Inverters},
  journal = {Reuters},
  year         = {2025},
  month        = {May 14},
  day          = {14},
  url          = {https://t.co/QhxNsVXxc7},
  note         = {Accessed: 2025-05-15}
}

@misc{helix_calc_tool,
  author       = {dereksgc},
  title        = {Helix Antenna Calculator},
  howpublished = {\url{https://sgcderek.github.io/tools/helix-calc.html}},
  note         = {Accessed: 2025-05-15},
  year         = {2024}
}

@book{kraus_antennas_2002,
	title = {Antennas for All Applications},
	isbn = {978-0-07-112240-5},
	pages = {968},
	publisher = {{McGraw}-Hill},
	author = {Kraus, John Daniel and Marhefka, Ronald J.},
	year = {2002},
    edition = {3},
	langid = {english},
    address = {New York, NY}
}

@misc{dsgc_helix_v5,
  author       = {dereksgc},
  title        = {dsgc\_helix\_v5.scad: Helix Antenna Scaffold},
  howpublished = {\url{https://github.com/sgcderek/helix-antenna-scaffold/blob/master/dsgc_helix_v5.scad}},
  note         = {Accessed: 2025-05-15},
  year         = {2024}
}

@misc{openscad,
  author       = {{OpenSCAD Developers}},
  title        = {{OpenSCAD}: The Programmers Solid {3D} {CAD} Modeller},
  howpublished = {\url{https://openscad.org}},
  note         = {Accessed: 2025-05-15},
  year         = {2024}
}

@misc{prusa_mk1,
  author       = {{Prusa Research}},
  title        = {Original Prusa i3 {MK1} {3D} Printer},
  howpublished = {\url{http://www.prusa3d.com/#our-printer}},
  note         = {Accessed: 2015-05-15},
  year         = {2015}
}

@misc{prusament_petg,
  author       = {{Prusa Polymers}},
  title        = {Prusament {PETG} Recycled},
  howpublished = {\url{https://prusament.com/materials/prusament-petg-recycled/}},
  note         = {Accessed: 2025-05-15},
  year         = {2022}
}

@INPROCEEDINGS{4584946,
  author={Olcan, Dragan I. and Zajic, Alenka R. and Ilic, Milan M. and Djordjevic, Antonije R.},
  booktitle={Proc. 1st Eur. Conf. Antennas Propag. (EuCAP)}, 
  title={On the optimal dimensions of helical antenna with truncated-cone reflector}, 
  year={2006},
  volume={},
  number={},
  pages={1-6},
  doi={10.1109/EUCAP.2006.4584946}}

@misc{sawbird_goes,
  author       = {{Nooelec LLC}},
  title        = {{SAWbird+} {GOES}: Premium {SAW} Filter \& Cascaded Ultra-Low Noise {LNA} Module},
  howpublished = {\url{https://www.nooelec.com/store/sawbird-plus-goes.html}},
  note         = {Accessed: 2025-05-15},
  year         = {2022}
}

@misc{thing6436342,
  author       = {t0nito},
  title        = {1.7 {GHz} {HRPT} Helicone Antenna},
  year         = {2024},
  howpublished = {\url{https://www.thingiverse.com/thing:6436342}},
  note         = {Accessed: 2025-05-16}
}

@misc{portapack_h4m,
  author       = {{Lab401}},
  title        = {{PortaPack {H4M} + {HackRF} All-in-one}},
  year         = {2025},
  howpublished = {\url{https://lab401.com/products/portapack-h4m}},
  note         = {Accessed: 2025-05-16}
}

@incollection{BalanisConstantineA2016AM,
author = {Balanis, Constantine A},
booktitle = {Antenna Theory: Analysis and Design},
copyright = {2016},
edition = {4},
isbn = {9781118642061},
language = {eng},
pages = {981-1026},
publisher = {John Wiley \& Sons},
title = {Antenna Measurements},
year = {2016},
}

@techreport{schmidt1973diffraction,
  author       = {Richard F. Schmidt},
  title        = {Diffraction Studies Applicable to 60-Foot Microwave Research Facilities},
  year         = {1973},
  number       = {NASA-TM-X-70707},
  institution  = {NASA Goddard Space Flight Center},
}

\end{document}